\documentclass[aps,twocolumn,superscriptaddress,floatfix,longbibliography]{revtex4-1}
\usepackage{tikz}
\usepackage{lipsum}
\usepackage{graphicx}
\usepackage[export]{adjustbox}
\usepackage{amsmath,amssymb,wasysym}
\usepackage{braket}
\usepackage{mathtools}
\usepackage{mathrsfs}
\usepackage{verbatim}
\usepackage{makecell}
\usepackage{cases}
\usepackage{tikz}
\usepackage[compat=1.1.0]{tikz-feynman}

\usepackage{slashed}
\usepackage{hyperref}
\usepackage{xcite}
\usepackage{xr}

\usepackage{xcolor}
\usepackage{pgfplots}
\pgfplotsset{compat=1.18}
\makeatletter
\newcommand{\im}{i}

\newcommand{\sgn}{\mathrm{sgn}}

\newcommand{\sech}{\mathrm{sech}}

\renewcommand{\Im}{\mathrm{Im}}
\renewcommand{\Re}{\mathrm{Re}}

\newcommand{\BigO}{\mathcal{O}}

\def\Q{\mathbf{Q}}

\def\k{\mathbf{k}}
\def\q{\mathbf{q}}
\def\x{\mathbf{x}}

\begin{document}
\title{Theory of Linear Magnetoresistance in a Strange Metal}

\author{Jaewon Kim}
\affiliation{Department of Physics, University of California, Berkeley, California 94720, USA}
\affiliation{Department of Physics and Anthony J. Leggett Institute for Condensed Matter Theory, University of Illinois Urbana-Champaign, Urbana, Illinois 61801, USA} 

\author{Shubhayu Chatterjee}
\affiliation{Department of Physics, Carnegie Mellon University, Pittsburgh, Pennsylvania 15213, USA}

\begin{abstract}
A central puzzle in strongly correlated electronic phases is strange metallic transport,  marked by $T$-linear resistivity and $B$-linear magnetoresistance, in sharp contrast with  quadratic scalings observed in conventional metals.
Here, we demonstrate that proximity to quantum critical points, a recurring motif in the phase diagrams of strange metal candidates, can explain both transport anomalies. 
We construct and solve a minimal microscopic model by coupling electronic excitations at the Fermi surface to quantum critical bosons via a spatially disordered Yukawa interaction, as well as static pinned domains of density wave order. 
The resultant transport relaxation rate scales as $k_B T/\hbar$ at low magnetic fields, and as an effective Bohr magneton $\tilde{\mu}_B B/\hbar$ at low temperatures.
Further, the magnetoresistance in our model shows a scaling collapse upon rescaling the magnetic field and the resistance by temperature, in agreement with experimental observations.
\end{abstract}
\maketitle

\textit{Introduction.--}
Strongly correlated metals often exhibit unconventional electronic transport that defy the basic tenets of semiclassical Boltzmann theory. 
One prototypical example of such behavior is the strange metal phase, observed ubiquitously across several (quasi) two-dimensional materials, ranging from high T$_c$ cuprates to heavy fermion compounds, and more recently, in moir\'e graphene \cite{LSCO,2009Sci...323..603C,2022Natur.601..205Y,Hayes,FeSe,FeSe2,Ghiotto,Jaoui,2022Natur.601..205Y}. 
Typically seen in proximity to symmetry-broken phases, the strange metal phase shows two distinctive anomalous features in its transport.
First, its zero-field resistivity $\rho(T)$ is $T$-linear down to low temperatures with a universal \textit{Planckian} relaxation rate $\tau_T^{-1} \simeq k_B T/\hbar$ \cite{Bruin804}.
Second, its low-temperature magnetoresistance $\rho(B)$ scales linearly with the magnetic field $B$ \cite{LSCO,2009Sci...323..603C,2022Natur.601..205Y,Hayes,FeSe,FeSe2,Ghiotto,Jaoui,2022Natur.601..205Y}, with an analogous relaxation rate set by the effective Bohr magneton, $\tau_B^{-1} \simeq \tilde{\mu}_B B/\hbar = e B/\tilde{m}_f$  ($\tilde{m}_f$ is the effective electronic mass) \cite{LMR}. 
Both these features are in stark contrast to 
semiclassical transport theory, which predicts $\rho(T) \sim T^2$ and $\rho(B) \sim B^2$ at low temperatures \cite{Ashcroft76,ziman1979principles,mah00}. 
The widespread observation of strange metallic transport in correlated metals implies physics beyond the traditional Landau paradigm of Fermi liquids with quasiparticle excitations \cite{Phillips,Greene,Chowdhury1,Chowdhury2,sachdev2025footfancupratephase,chang2024,grilli2022,Aldape:2020enq,Patel2,Patel3,Kim2024,esterlis2021large,Bashan2024,bashan2025,Evyatar2024}, and calls for a minimal microscopic model that simultaneously accounts for the two distinct aspects of unconventional transport.

In this Letter, we provide a unifying explanation for the origin of both transport anomalies by leveraging a common feature in the phase diagram of strange metals---proximity to quantum critical points with symmetry-breaking orders.
Specifically, we construct a simple microscopic model of electrons coupled to nearly critical order parameters with spatial disorder, and calculate its resistivity $\rho(B,T)$. 
Within our model, the dynamical coupling between the electrons at the Fermi surface and critical bosons leads to non-Fermi liquid behavior with $T$-linear resistivity and Planckian dissipation \cite{Aldape:2020enq,Patel2,Patel3}, while a static coupling to pinned order parameter domains induces $B$-linear magnetoresistance (LMR) with a universal slope \cite{LMR}. 

\begin{figure}
    \centering
    \includegraphics[width = 1.0\columnwidth]{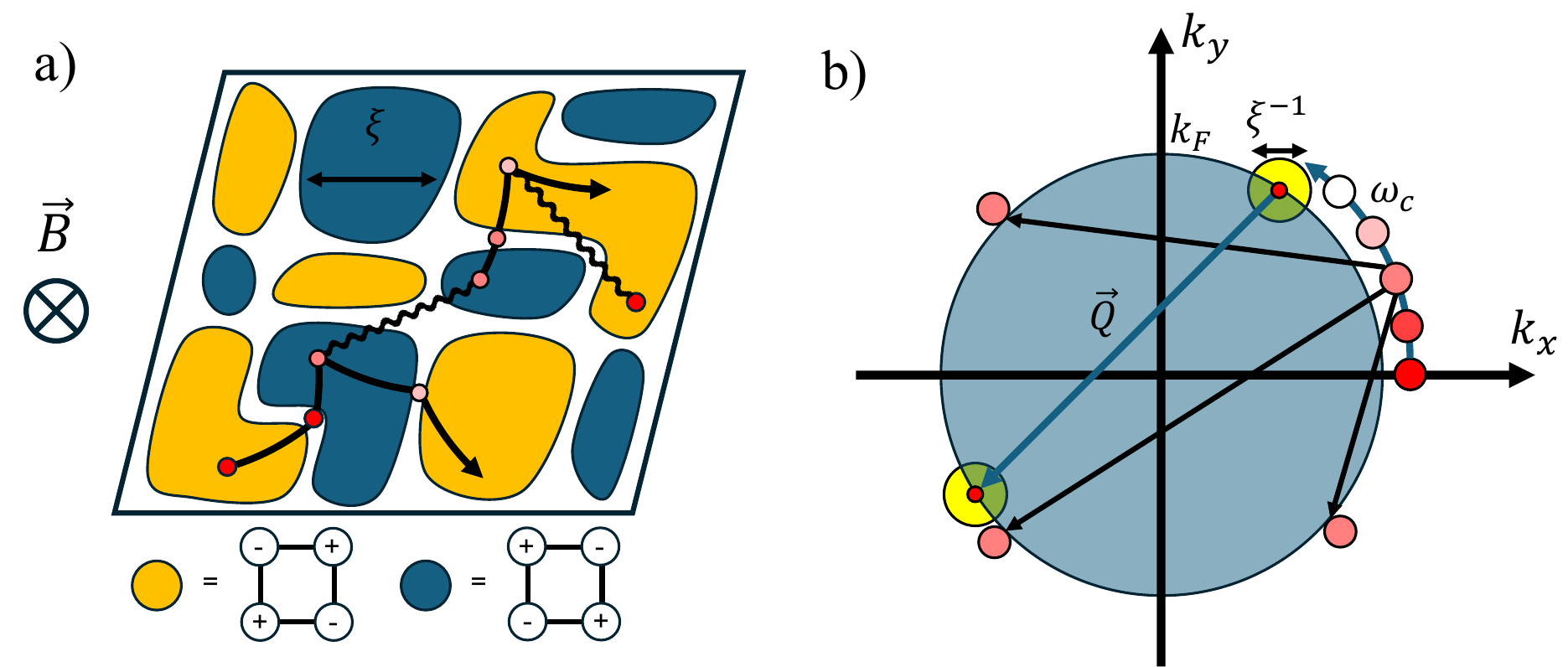}
    \caption{(a) 
    A real space schematic of our model, with electrons moving in the vicinity of a critical point with disorder-pinned domains of density wave order.
    Inelastic scattering from critical bosons (denoted by squiggly lines) gives rise to a marginal Fermi liquid, and elastic scattering from glassy density waves lead to LMR.
    (b) Momentum space schematic: a magnetic field rotates excitations at the Fermi surface into hot spots of size $\xi^{-1}$ (yellow circles) where they incoherently backscatter and relax momenta, resulting in an overall relaxation rate of $\BigO(\omega_c)$ and a $B$-linear magnetoresistance.
    The excitations may also decay through collisions with critical bosons, as depicted via the gradual color change from red to white.
    The latter process relaxes the momentum of an excitation over the entire Fermi surface and gives rise to $T$-linear resistivity.}
    \label{fig:headimage}
\end{figure}

Our main results are threefold.
First, by framing and solving a quantum Boltzmann equation for our microscopic model, we explicitly show that the resistivity $\rho(B,T)$ is $T$-linear at low magnetic fields and $B$-linear at low temperatures, with the desired universal slopes in both cases.
Second, we obtain a scaling collapse upon rescaling the magnetoresistance and the magnetic field by temperature, in agreement with experimental observations in several correlated metals \cite{LSCO,2022Natur.601..205Y,2009Sci...323..603C,FeSe,FeSe2,Hayes,Ghiotto}, and additionally determine an explicit analytical form of the scaling function.
Third, we numerically establish a concrete lower bound on the magnetic field for observing LMR, and provide an intuitive argument to justify such a bound even in the absence of well-defined quasiparticles at the Fermi surface. 
Collectively, our results demonstrate that the universal transport phenomenology of strange metals can be obtained from simple microscopic ingredients that are omnipresent in the phase diagrams of correlated quantum materials.

\textit{Model.--}
In the vicinity of quantum critical points, low-energy fermions at the Fermi surface couple to dynamical order parameter fluctuations, as well as static pinned domains with large correlation lengths. 
To capture the destruction of quasiparticles in the presence of disorder, we consider a local, spatially disordered Yukawa coupling between the critical bosons ($\phi$) and spinless fermions ($c$).
Recent seminal work \cite{Aldape:2020enq,Patel2,Patel3,esterlis2021large} has shown that such a coupling, upon disorder averaging, leads to a marginal Fermi liquid (mFL) phase---characterized by a sharp Fermi surface that separates occupied from unoccupied states in momentum space, but lacking well-defined quasiparticles at the Fermi surface as the low-energy excitations are severely short-lived with a divergent decay rate \cite{Varma}.
We additionally include coupling of fermions to static charge density-wave domains of typical size $\xi$ \cite{LMR}.
Taken together, the Hamiltonian of our minimal microscopic model is given by \footnote{Henceforth, we set $\hbar = 1$. But we will put $\hbar$ back in while discussing the universal slopes of scattering rates},
\begin{align}
H &= \underbrace{H_f + H_b +  H_{bf}}_{H_{\rm mFL}} + H_{\rm dw} \nonumber \\
 H_f &= \int_\k \left( \frac{k^2}{2m_f} - \mu \right)c_\k^\dagger c_\k, ~ H_b = \int_\q \left(m_b^2 + q^2 \right) \phi_\q \phi_{-\q}, \, \nonumber \\
 H_{bf} &= \sum_\x g_\x \phi_\x c^\dagger_\x c_\x
    \textrm{ where } \overline{g_\x} = 0, \ \overline{g_\x g_{\x'}} = g^2 \delta_{\x \x'}, \nonumber \\
 H_{\textrm{dw}} &= J \sum_\x n_\x e^{i \Q \cdot \x} c^\dagger_\x c_\x \, , \,  \overline{n_\x} = 0, \,  \overline{n_\x n_{\x'}} = e^{-\frac{|\x-\x'|^2}{4\xi^2}}. 
\label{eq:H}
\end{align}
In Eq.~\eqref{eq:H}, $H_f$ is the free-fermionic Hamiltonian (assumed quadratic for simplicity) with Fermi energy $\varepsilon_F$, and $H_b$ denotes the bosonic Hamiltonian with a boson mass $m_b$ that goes to zero at criticality. 
$H_{bf}$ denotes the disordered Yukawa coupling $g_\x$ between the fermions and the dynamically fluctuating critical bosons: such coupling disorder in $g_\x$ is expected to originate from local variations in hopping ($t_{ij}$) or interaction (Hubbard $U_i$) \cite{Patel3,ZhangBultinck2025}. 
Finally, $H_{\rm dw}$ denotes the coupling of fermions to glassy charge density-wave order (CDW) at momentum $\Q$, with the CDW amplitude $n_\x$ being correlated over length scales of $\xi$ much larger than the microscopic lattice spacing. 
Such glassy density-wave orders arise naturally from the simultaneous presence of strong electronic correlations that promote symmetry-breaking orders~\cite{FradkinKivelsonTranquada_RMP_intertwined,KivelsonRMP_stripes,fujita2014direct,Hamidian_2015,laplaca,lu2017short,Curro_pnictide,regan2020mott,zong2025quantum,jin2021stripe}, and quenched disorder in the form of structural inhomogeneity~\cite{pan2001microscopic}, random dopant distribution~\cite{Kato2005} or local strains~\cite{lau2022reproducibility} that can pin density-wave domains~\cite{KivelsonRMP_stripes}.

\begin{figure}
    \centering
    \includegraphics[width = 1\columnwidth]{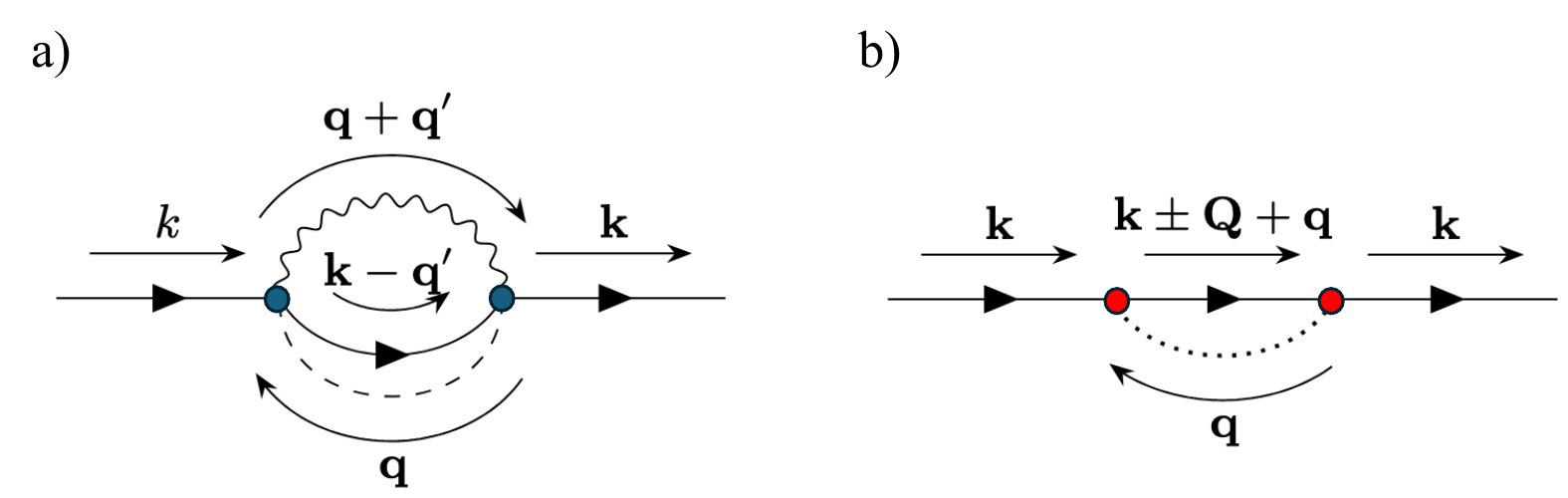}
    \caption{Electron self-energy diagrams. (a) Self-energy $\Sigma_{\rm mFL}$ due to the spatially random Yukawa interactions in $H_{bf}$. The solid lines denote electrons, the squiggly line denotes the critical bosons, and the dashed line corresponds to disorder averaging over the Yukawa vertex $g_{\x}$ (blue).  (b) Self-energy $\Sigma_{\rm dw}$ due to elastic scattering off glassy density waves in $H_{\rm dw}$. The dotted line indicates disorder-averaging over the static density-wave vertex $n_\x$ (red).}
    \label{fig:SelfE}
\end{figure}

\textit{Self-energy.--}
We now analyze the effect of each fermion-boson  coupling in turn. 
This can be succinctly captured via the self-energy correction $\Sigma(\k,\omega)$ to the bare fermionic Green's function $G_0(\k,\omega)$.
\begin{align}
G^{-1}(\k, \omega) &= G_{0}^{-1}(\k,\omega) - \Sigma(\k, \omega), \text{ where } \nonumber \\
G_{0}^{-1}(\k,\omega) &= \omega + i \delta - \left( \frac{k^2}{2m_f} - \mu \right) 
\label{eq:G}
\end{align}
The self-energy receives contributions from both $H_{bf}$ and $H_{\rm dw}$, i.e., $\Sigma(\k,\omega) = \Sigma_{\rm mFL}(\k,\omega) + \Sigma_{\rm dw}(\k,\omega)$, where each part is given by the corresponding Feynman diagram in Fig.~\ref{fig:SelfE}.
While such a one-loop self-energy calculation is inherently perturbative, our results are exact in a generalization of our model with a large number of fermion and boson flavors (large $N$ limit).
Delegating the details of the large $N$ model to the Supplemental Material~\cite{SM}, here we focus on the important physical features of $\Sigma(\k,\omega)$ and their implications for transport.

For the boson-fermion coupling $H_{bf}$, we set the average Yukawa coupling $\bar{g} = 0$, as the disordered part of the interaction leads to momentum relaxation and determines the transport lifetime \cite{Patel3}. 
As the bosons are tuned toward criticality ($m_b \to 0$), the fermionic self-energy takes an mFL form, as found in  Refs.~\cite{Aldape:2020enq,Patel2,Patel3,Kim2024,esterlis2021large} for closely related models.
\begin{equation}
    \Sigma_{\rm mFL}(\k = k_F \hat{k}, \omega) \simeq -\frac{m_f g^2 \omega}{4\pi^2} \log\frac{\Lambda}{m_T^2 - \frac{m_f^2 g^2}{4\pi} i \omega} \,,
    \label{eq:Sigmaint}
\end{equation}
where $m_T^2 \sim m_f^2 g^2 T$ denotes a thermal boson mass that opens up at finite temperatures $T$, and $\Lambda$ is an appropriate ultra-violet cutoff \cite{Podolsky,Aldape:2020enq,Patel2,Patel3,Kim2024,esterlis2021large}.

Eq.~\eqref{eq:Sigmaint} has two important consequences for transport. 
First, it implies a renormalization of the quasiparticle weight $Z$, and consequently, a renormalization of the effective quasiparticle mass to $\tilde m_f = m_f/Z$, where 
\begin{equation}
    Z \simeq \left( 1 + \frac{m_f g^2}{4\pi^2} \log \frac{\Lambda}{T} \right)^{-1} \,.
    \label{eq:Z}
\end{equation}
Second, the dissipation rate for fermions scattering with the critical bosons can be inferred from Eq.~\eqref{eq:Sigmaint}, as the vertex correction vanishes due to the isotropic momentum-independent nature of the vertex \cite{Patel3}.
The resultant momentum relaxation rate $\tau_{\rm re}^{-1}$ of the fermions, which can be extracted from the Drude formula as $\tau_{\rm re}^{-1} = (n e^2 \rho)/\tilde{m}_f$, satisfies
\begin{equation}
    \tau_{\rm re}^{-1} \simeq \alpha \left( \frac{k_B T}{\hbar} \right) \,, \textrm{ where } \alpha \simeq \frac{\frac{m_f g^2}{4\pi^2} \log \frac{\Lambda}{T}}{1 + \frac{m_f g^2}{4\pi^2} \log \frac{\Lambda}{T}}\,.
    \label{eq:taure}
\end{equation}
In the strong coupling limit $m_f g^2 \gg 1$, $\alpha \to 1$, and the relaxation rate becomes \textit{Planckian}, $\tau_{\rm re}^{-1} \to \tau_T^{-1} \simeq k_B T/\hbar$.

\begin{figure*}[htbp]
    \centering
    \includegraphics[width = \textwidth]{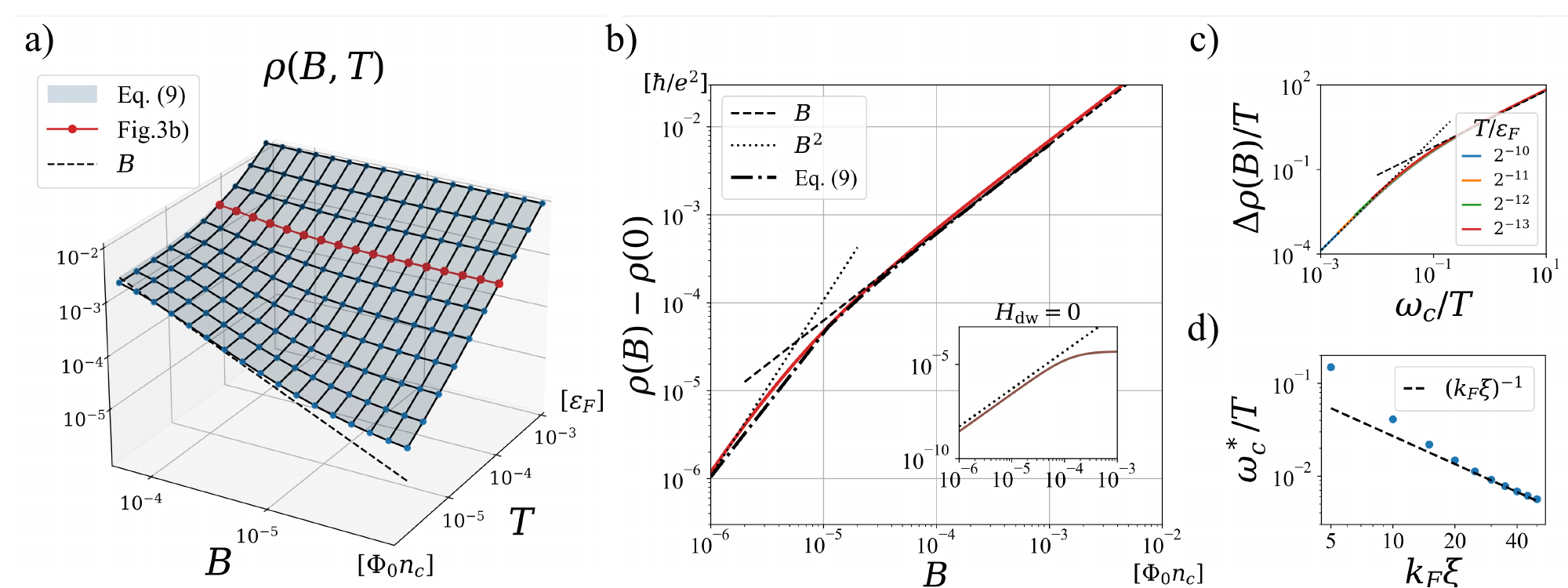}
    \caption{(a) The resistivity $\rho(B,T)$ as a function of temperature ($T$) and the magnetic field (parametrized by $\omega_c$), obtained by numerically solving the quantum Boltzmann equation (dots) for $k_F \xi = 20$, $J = 0.05 \, \varepsilon_F$ and $m_f g^2 = 4$ [$\hbar = 1 = k_B$].
    Both $T$-linear resistivity at low fields, and $B$-linear magnetoresistance at low temperatures are apparent. 
    An excellent agreement is obtained with the analytical scaling function (light blue surface) in Eq.~\eqref{eq:result}.
    (b) A line cut of panel (a) at a fixed temperature $T = 2^{-12} \varepsilon_F$ (red line), showing the magnetoresistance $\Delta \rho(B) \equiv \rho(B,T) - \rho(0,T)$.
    $\Delta \rho(B)$ initially scales quadratically with the magnetic field (dotted line), but crosses over to scaling linearly at $\omega_c^*$ where the dotted  line ($\propto B^2$) and the dashed line ($\propto B$) cross. 
    In the linear regime, $\Delta \rho(B) \simeq \tilde{m}_f/(n_c e^2 \tau_B)$ (dashed line), where $\tau_B^{-1} = \tilde{\mu}_B B/\hbar$.
    In the entire magnetic field range, the magnetoresistance is well-captured by our ansatz in Eq.\eqref{eq:result} (dash-dotted line). 
    Inset: solution to the QBE with $H_{\rm dw} = 0$ at the same $T$, showing $\Delta \rho (B)_{\rm mFL} \propto B^2$ (dotted line) before it saturates. 
    The marginal Fermi liquid state does not exhibit $B$-linear magnetoresistance without glassy density-wave order.
    (c) Scaling collapse of $\Delta \rho(B)$ when both the magnetoresistance and the magnetic field are rescaled by the temperature $T$.
    Legend indicates the different temperatures, $\beta\varepsilon_F$. 
    (d) Numerical evidence for power law scaling of $\omega_c^*/T$ with $k_F\xi$.
    }
    \label{fig:MainResults}
\end{figure*}

For the CDW-fermion coupling $H_{\rm dw}$, we consider the weak coupling limit $J \lesssim v_F/\xi$ so that a description in terms of a single large Fermi surface suffices \cite{LMR}. 
In this limit, the fermionic self-energy due to scattering from glassy CDW order takes the form 
\begin{equation}
    \Sigma_{\textrm{dw}}(\k, \omega) = \int_{\q} J^2 \xi^2 e^{-\xi^2 q^2} G(\k \pm \Q + \q,\omega)
    \label{eq:SigmaCDW}
\end{equation}
where $G(\k,\omega)$ denotes the fermion Green's function.
Physically, this indicates that electrons incoherently backscatter between the hot spots with a large momentum transfer $\approx \Q$, efficiently relaxing momentum (Fig.~\ref{fig:headimage}).
Furthermore, the lack of long-range CDW order, as exemplified by a finite correlation length $\xi$ in Eq.~\eqref{eq:H}, smears the hot spots by a momentum scale $\BigO(\xi^{-1})$.
This means that hot regions occupy only a small angular extent of $\BigO(1/k_F\xi)$ of the Fermi surface.
Ergo, in the limit of large $k_F\xi$, the coupling to glassy density waves leads to a significant modification of the self-energy near these isolated hot-regions, where $\Sigma(\k,\omega) \approx \Sigma_{\rm dw}(\k,\omega) \approx (k_F \xi)J^2/\varepsilon_F$~\cite{LMR}\footnote{To obtain this estimate of the magnitude of the self-energy at the hot spots, we approximate $\Sigma \approx \Sigma_{\rm mFL}$ and evaluate $\Sigma_{\rm dw}$ using Eq.~(6). This result coincides with the expectation from a semiclassical treatment that assumes well-defined quasiparticles in Ref.~\onlinecite{LMR}. Although quasiparticles are no longer sharply defined in the mFL, the fermionic spectral weight remains strongly concentrated near the Fermi surface, with a width of order $T$. Consequently, when $T \ll v_F / \xi$, density-wave disorder is insensitive to whether quasiparticles are well defined, and the resulting self-energy retains the same functional form.}. 
However, such hot spot scattering does not significantly modify $\Sigma(\k,\omega)$ away from the hot spots, i.e., $\Sigma(\k,\omega) \approx \Sigma_{\rm mFL}(\k,\omega)$ for most of the Fermi surface.
Thus, our microscopic model yields an mFL, with additional elastic scattering near isolated hot regions on the Fermi surface.

\textit{Transport properties.--}
To find the dc electrical resistivity of our model, we turn to the quantum Boltzmann equation (QBE), which models the transport in terms of a generalized distribution function despite the absence of sharply defined quasiparticles \cite{PrangeKadanoff,KimLeeWen,NaveLee}. 
Specifically, the divergent self-energy $\Sigma_{\rm mFL}$ for fermions at the Fermi surface indicates a breakdown of fermionic quasiparticle excitations, typically used to derive a semiclassical Boltzmann equation \cite{mah00}.
Nevertheless, it is possible to formulate a generalized QBE provided the fermionic self-energy $\Sigma(\k,\omega)$ is independent of the magnitude of momentum $\k$ near the Fermi surface, and the spectral function $A(\k,\omega)$, expressed in terms of the energy $\xi_{\k} = k^2/2m_f - \mu$, remains peaked at $\xi_k = 0$ for small $\omega$. 
These features, which our model shares with 
transport problems involving strong coupling of fermionic excitations with phonons \cite{PrangeKadanoff} or emergent gauge fields \cite{KimLeeWen,NaveLee}, allow us to analogously derive the following QBE.
\begin{align}
   & [A(\k,\omega)]^2 \, \Im[\Sigma^R(\k,\omega)] \, \left( \frac{e \k \cdot \mathbf{E}}{m_f} \right) {f_0}'(\omega) \nonumber \\
   & + \left(\frac{e \k \times B\hat{z}}{m_f} \right) \cdot \nabla_{\k} G^<(\k,\omega) = \Sigma^> G^< - \Sigma^< G^> \,,
    \label{eq:BEf}
\end{align}
where $f_0(\omega) = (1 + e^{\beta \omega})^{-1}$ is the equilibrium Fermi-Dirac distribution, $G^{>,<}$ denote the greater and lesser Green's functions and $\Sigma^{>,<}$, their corresponding self-energies [suppressing $(\k,\omega)$ indices for clarity].
The left-hand side of Eq.~\eqref{eq:BEf} stands for the electromagnetic force on the fermions; and the right for the collision integrals for the scattering processes \cite{mah00,KimLeeWen,NaveLee}.
This QBE takes the fermionic self-energy $\Sigma(\k,\omega)$ computed previously as input, and describes the evolution of the generalized fermion distribution function in both momentum and frequency space in response to the external electromagnetic forces.
By numerically solving for the deviation of the fermionic distribution function from equilibrium due to the presence of electromagnetic fields, we obtain the current and subsequently the resistivity $\rho(B,T)$ \cite{SM}.

We present the results for our numerical solution of the QBE, choosing $\Q = (\pi,\pi)$ for concreteness, in Fig.~\ref{fig:MainResults}(a), where we plot $\rho(B,T)$ as a function of both magnetic field and temperature. 
At small magnetic fields, the resistivity is $T$-linear with a Planckian dissipation rate, consistent with our expectations that small hot-regions on the Fermi surface do not play a significant role in momentum relaxation, which is dominated by inelastic scattering from critical bosons \cite{Aldape:2020enq,Patel2,Patel3}. 
In this weak-field regime, the magnetoresistance $\Delta \rho(B) = \rho(B, T) - \rho(0,T)$ scales as $B^2$.
However, at larger magnetic fields, we note that there is a crossover to $B$-linear behavior in the magnetoresistance, i.e., 
\begin{equation}
\Delta \rho(B) \simeq \frac{\tilde{m}_f}{n_c e^2}\left(\frac{\tilde{\mu}_B B}{\hbar}\right)    
\end{equation}
with a universal slope $\tau_B^{-1} \simeq  \tilde{\mu}_B B/\hbar = eB/\tilde{m}_f = \omega_c$.
To see this behavior of $\Delta \rho(B)$ more explicitly, we plot a line cut in Fig.~\ref{fig:MainResults}(b) at a fixed small temperature, where a pronounced $B$-linear regime is apparent.  
Further, the numerical data for the magnetoresistance shows a scaling collapse when both axes---$\Delta \rho(B)$ and $\omega_c$ are rescaled by the temperature $T$, which sets the zero-field resistance (Fig.~\ref{fig:MainResults}(c)).
Such scaling behavior---an analog of Kohler's rule in Fermi liquids \cite{kohler1938magnetischen}, is consistent with experiments across many strange metal candidates \cite{LSCO,2009Sci...323..603C,2022Natur.601..205Y,Hayes,FeSe,Ghiotto,2022Natur.601..205Y}. 
By contrast, if we explicitly set $H_{\rm dw} = 0$, our solution to the QBE for $H_{\rm mFL}$ gives a magnetoresistance $\Delta \rho(B)_{\rm mFL} \propto B^2$ until it saturates at higher fields, with no intermediate $B$-linear regime (Fig.~\ref{fig:MainResults}(b), inset).
Therefore, we conclude that the additional coupling between fermions and glassy density-wave order is crucial for the \textit{simultaneous observation} of $B$-linearity and $B/T$ scaling of the magnetoresistance in a marginal Fermi liquid.  

An intuitive understanding of the origin of LMR, as well as the crossover magnetic field scale from Fermi-liquid-like ($B^2$) to $B$-linear behavior, may be obtained via a semiclassical picture \cite{LMR}. 
When the hot spot scattering rate is large~\cite{SM}, 
the dominant momentum-relaxation mechanism for excitations at the Fermi surface is to rotate into the hot spot regions and then incoherently backscatter off the glassy density-wave order. 
Thus, the overall relaxation rate is set by the cyclotron frequency which sets how fast an excitation can rotate into the hot region, i.e., $\tau_B^{-1} \simeq \omega_c = eB/\tilde{m}_f = \tilde{\mu}_B B$.
Further, for elastic scattering from glassy density waves to dominate momentum relaxation, a fermionic excitation at the fringe of the hot region should rotate fast enough to avoid decay via emission of critical bosons. 
Specifically, if the time required to rotate by an angle $\Delta \theta = 1/(k_F \xi)$ on the Fermi surface, given by $\Delta t = \Delta \theta/\omega_c$, is larger than the mFL lifetime $\tau_T$, an electronic excitation will relax via emission of critical bosons, and backscattering from hot spots is rendered ineffective.  
Consequently, we expect to see LMR beyond a minimum cyclotron frequency $\omega_c^* = \Delta \theta/\tau_T = k_B T/(k_F \xi)$. 
This is borne out by our numerical data in Fig.~\ref{fig:MainResults}(d), where we extract the crossover frequency $\omega_c^*$ as a function of density-wave correlation length $\xi$, and show that it scales as $T/k_F \xi$.

Collectively, our observations suggest the following scaling form for the magnetoresistance. 
\begin{equation}
    \rho(B,T) = \frac{\tilde{m}_f}{n_c e^2} \left[ \left(\alpha-\gamma \right)k_B T + \sqrt{(\tilde{\mu}_B B)^2 + \gamma^2 (k_B T)^2 }  \right] \,,
    \label{eq:result}
\end{equation}
where $\alpha$ denotes the Planckian coefficient from Eq.\eqref{eq:taure}, and $\gamma \simeq \alpha/(k_F\xi)$ determines the crossover field at which LMR sets in via $\omega_c^* = \tilde{\mu}_B B^* \simeq \gamma k_B T \sim k_B T/(k_F\xi)$.
In Fig.~\ref{fig:MainResults}(a), we compare our numerical results for $\rho(B,T)$ with the ansatz in Eq.~\eqref{eq:result}, finding an excellent agreement over the entire range of magnetic field and temperature considered.

\textit{Summary and Discussion.-}
In this Letter, we demonstrated that two distinct ubiquitous aspects of strange metallic transport, $T$-linear zero-field resistivity with Planckian dissipation and $B$-linear low-temperature magnetoresistance with a universal relaxation rate, can simultaneously arise from proximity to quantum critical points. 
While we considered spinless fermions coupled to charge density waves for simplicity, our theoretical framework is readily adaptable to other density-wave orderings, such as bond-density waves \cite{fujita2014direct,Hamidian_2015,laplaca} or spin-density waves for spinful fermions \cite{AuerbachBook}. 
More generally, going beyond our specific model, our work establishes unconventional magnetotransport when excitations at a Fermi surface are coupled to static glassy density wave order with large domains, regardless of the presence of well-defined quasiparticles.  
If the low-energy physics is that of a marginal Fermi liquid, we expect to see strange metallic transport; otherwise, the coupling $H_{\rm dw}$ leads to LMR in a Fermi liquid with its characteristic $T^2$ zero-field resistivity \cite{LMR}. 
Therefore, coupling to glassy density waves is a very general mechanism of LMR across a variety of correlated metals \cite{yu2021unusual}, independent of the temperature scaling of resistivity.

The magnetoresistance in our model has a $B/T$ scaling collapse, in alignment with experimental observations, albeit with a slight deviation from the proposed quadrature scaling $\rho(B, T) \propto \sqrt{(\mu_B B)^2 + (k_B T)^2}$ in Refs.~\onlinecite{Hayes,FeSe}.
Further, the magnetoresistance in our model for $\omega_c \gtrsim \omega_c^*$ scales as $\rho(B,T) \propto (\alpha - \gamma)k_B T + \tilde{\mu}_B B $, in accordance with the measured scaling $0.5 \, k_B T + \mu_B B$ in recent experiments on nanopatterned YBCO \cite{2022Natur.601..205Y}.
Finally, we estimate a crossover field scale $B^*/T = 0.5$ Tesla per Kelvin for moderate disorder strength $k_F \xi \approx 10$, and $m_f = 4 m_e$ \cite{LSCOmass2}, in reasonable agreement with experiments on cuprates and pnictides \cite{Hayes,LSCO}. 

In the literature, random resistor networks~\cite{Parish,PL2005} and effective medium theories~\cite{Stroud75,GuttalStroud2005,Adam2017} have been used to phenomenologically model LMR arising from variation of carrier density due to macroscopic disorder in Fermi liquids.
Here, the net longitudinal resistance depends on local Hall resistances that scale linearly in $B$, as the disorder leads to distorted current paths perpendicular to the global electric field~\cite{Parish}. 
Ref.~\cite{Aavishkar} applied the effective medium picture to a mFL, and obtained a crossover scale $\omega_c^*$ that depends only on the temperature $T$.  
By contrast, LMR is obtained by directly solving for transport in our microscopic electronic Hamiltonian at fixed carrier density.
Further, the crossover scale $\omega_c^* \sim k_B T/k_F \xi$ depends on both $T$ and the typical domain size $\xi$ of pinned density waves, and can vary across materials.

Our results are valid at temperatures above any potential superconducting instability of $H_{\rm mFL}$, expected for real (but not complex) Yukawa couplings \cite{Li_2024}.
Further, we have neglected any potential short-range disorder scattering, as well as the regime of disorder induced localization of the critical bosons, observed in recent numerical studies of $H_{\rm mFL}$ \cite{Patel_2024}.
While the former is expected to not modify the transport properties beyond an innocuous residual resistivity at $T = 0$ and can hence be neglected on account of weak residual resistivity of most strange metals, the occurrence of boson localization in a magnetic field and its effect on magnetotransport remain open questions. 
Additionally, our work also sets the stage to study magnetotransport in other kinds of non-Fermi liquids, e.g., those obtained via coupling disordered two-level-systems to fermions \cite{Evyatar2024,Bashan2024}, or arising from mesoscale superconducting puddles \cite{bashan2025}.

Finally, analogous to glassy density waves, we showed in Ref.~\onlinecite{LMR} that glassy nodal order, such as nematic order on a square lattice, can also lead to LMR in a conventional Fermi liquid. 
In this case, the presence of cold spots on the Fermi surface creates a bottleneck for momentum relaxation, which is removed by a magnetic field that rotates the fermions across these cold regions, leading to a momentum relaxation rate set by the cyclotron frequency $\omega_c$ and consequently, LMR.
From this physical picture, we expect that a resistance $\rho(B,T)$ that is both $T$-linear and $B$-linear can also be obtained when a marginal Fermi liquid is coupled to glassy nodal order, albeit without the Kohler-like $B/T$ scaling. Further, since the mechanism is reliant on the presence of a momentum relaxation bottleneck, the LMR regime may be pushed down to low temperatures or even completely wiped out if the inelastic scattering off the critical bosons relaxes momentum faster than elastic scattering from glassy nodal order. 
We leave an explicit study of magnetotransport due to coupling between a marginal Fermi liquid and other kinds of symmetry-breaking orders, such as glassy nodal order, to future work. 

\textit{Acknowledgements.-} We gratefully acknowledge a collaboration with E. Altman on a related topic \cite{LMR}.
We also thank E. Berg, D. Chowdhury, C. H. Chung and T. Cookmeyer for insightful discussions.
J. K. acknowledges support from the National Science Foundation under Grant No. DMR-2225920.

\bibliography{PDLMR.bib}

\newpage

\onecolumngrid

\vspace{0.3cm}

\title{Supplemental Material: Theory of Linear Magnetoresistance in a Strange Metal}

\author{Jaewon Kim}
\affiliation{Department of Physics, University of California, Berkeley, CA 94720, USA}
\affiliation{Department of Physics and Anthony J. Leggett Institute for Condensed Matter Theory, University of Illinois Urbana-Champaign, Urbana, Illinois 61801, USA} 

\author{Shubhayu Chatterjee}
\affiliation{Department of Physics, Carnegie Mellon University, Pittsburgh, PA 15213, USA}

\maketitle

\begin{center}
\large{\bf{Supplemental Material: Theory of Linear Magnetoresistance in a Strange Metal}}
\end{center}

\section{Large $N$ Limit}
In this appendix, we present the large $N$ limit at which our results become analytically exact.
Recall that our Hamiltonian is given as, $H = H_{\textrm{mFL}} + H_{\textrm{dw}}$, where $H_{\textrm{mFL}}$ denotes the Hamiltonian of the marginal Fermi liquid, and $H_{\textrm{dw}}$, the coupling to the CDW disorder.
Following Ref.\cite{Aldape:2020enq,Patel2,Patel3}, the large $N$ Hamiltonian of the marginal Fermi liquid is given by,
\begin{equation}
\begin{split}
    & H_{\textrm{mFL}} = H_f + H_b + H_{bf} \,, \textrm{ where } \\
    & \quad H_f = \sum_{i=1}^N \int_k (\epsilon_k-\mu) c_{\k,i}^\dagger c_{\k,i} \,, \ \ H_b = \sum_{n=1}^N \int_k (m_b^2 + q^2) \phi_{q,n} \phi_{-q,n} \,, \ \ H_{bf} = \sum_{ij,n,\x} g_{\x,ij}^n \phi_{\x,n} c^\dagger_{\x,i} c_{\x,j} + c.c.
\end{split}
\end{equation}
where $g$ are gaussian random variables of zero mean with variance,
\begin{equation}
    \overline{g_{\x,ij}^n g_{\x',i'j'}^{n'}} = \frac{g^2}{N^2}\delta_{\x\x'} \delta_{ii'} \delta_{jj'} \delta_{nn'}\,.
\end{equation}

On the other hand, the coupling to the CDW disorder is given as $H_{\textrm{dw}}$,
\begin{equation}
    H_{\textrm{dw}} = n_{\x,ij} c^\dagger_{\x,i} c_{\x,j}
\end{equation}
where $n$ are also random variables of zero mean whose variance satisfies,
\begin{equation}
    \overline{n_{\x,ij} n_{\x',i'j'}} = (-1)^{\x-\x'} e^{-\frac{|\x-\x'|^2}{4\xi^2}} \frac{1}{2} \Big\{ \delta_{ii'}\delta_{jj'} + \delta_{ij'}\delta_{ji'} \Big\}
\end{equation}

The large $N$ saddle point is given by the following set of Schwinger-Dyson equations,
\begin{equation}
\begin{split}
    & G(\k,\omega) = \frac{1}{\im\omega - v_F (|k| - k_F) - \Sigma(\k,\omega)} \,, \quad \Sigma(\k,\omega) = \Sigma_{\rm mFL}(\omega) + \Sigma_{\rm dw}(\k,\omega) \\
    &\quad \Sigma_{\rm mFL}(\x, \tau) = g^2 \delta_{\x = 0} G(\x, \tau) \Big( F(\x, \tau) + F(-\x, -\tau) \Big) \\
    &\quad \Sigma_{\rm dw}(\k,\omega) = \int_q J^2 \xi^2 e^{-\xi^2 q^2} G(\k \pm \Q + \q, \omega) \\
    & F(\q, \Omega) = \frac{1}{m_b^2 + q^2 + \Pi(\Omega)} \,, \quad \Pi(\x, \tau) = g^2 \delta_{\x = 0} G(\x, \tau) G(-\x, -\tau) \,.
    \label{eq:appSD}
\end{split}
\end{equation}
where the fermion and boson Green's functions $G, F$ are defined as,
$$G(\x-\x', \tau-\tau') = \braket{c(\x, \tau) c^\dagger(\x', \tau')}\,, \ \ F(\x-\x', \tau-\tau') = \braket{\phi(\x,\tau) \phi(\x',\tau')}\,,$$
and $\Sigma$ and $\Pi$ denote their corresponding self-energies.

\subsection{Calculation of the Interaction Self-Energy}
Let us first derive $\Sigma_{\rm mFL}$, the fermion self-Energy due to the interaction with the critical bosons.
Foremost, we note that $\Sigma_{\rm mFL}$ is independent of the momentum.
This is due to the disorder in the interaction $g_\x$ being uncorrelated, which allows it to absorb any momentum.
Integrating $G(\k,\omega)$ with regards to $\k$, we find $G(\x = 0, \omega) = -\frac{\im m_f}{2}\sgn(\omega)$. Upon a convolution of $G(\x = 0, \omega)$ with itself in frequency, we obtain the boson self-energy, $\Pi$, and we find,
\begin{equation}
    \Pi(\Omega) = \frac{m_f^2 g^2}{4} \int_{\omega} \sgn(\omega+\Omega) \sgn(\omega) \simeq \Pi(0) + \alpha |\Omega| \,, \textrm{ where } \alpha =  \frac{m_f^2 g^2}{4\pi} \,.
    \label{eq:appPi}
\end{equation}

Therefore at the critical point, the boson propagator takes the following form,
\begin{equation}
    F(\q, \Omega) = \frac{1}{q^2 + \alpha \Omega}\,.
    \label{eq:appF}
\end{equation}
Integrating out $\q$, we find the $F(\x = 0, \Omega) = \frac{1}{4\pi}\log\frac{\Lambda_b}{\alpha |\Omega|}$.
Last, performing a convolution between $F(\x = 0)$ and $G(\x = 0)$, we find the fermion self-energy, which is given as,
\begin{equation}
    \Sigma(\Omega) = -2\frac{\im m_f g^2}{4\pi} \int_\Omega \sgn(\omega+\Omega) \log \frac{\Lambda}{\alpha |\Omega|}
    \simeq \frac{\im m_f g^2}{4\pi^2} \Omega \log \frac{\Lambda}{\alpha |\Omega|} \,.
    \label{eq:appSigma}
\end{equation}

At finite temperatures a thermal gap $m_T^2 \sim \alpha T$ opens up in the bosons \cite{Aldape:2020enq,Patel2,Podolsky,Kim2024}. In turn, this thermal gap modifies \eqref{eq:appSigma} to,
\begin{equation}
    \Sigma(\omega_n) \simeq -\frac{\im m_f g^2}{2\pi^2} \omega_n \log \frac{\Lambda_b}{m_T^2 + \alpha |\omega_n|} \,.
    \label{eq:appSigma_T}
\end{equation}

Analytically continuing \eqref{eq:appSigma_T} to real time, we find for small frequencies $\omega \ll m_T^2$,
\begin{equation}
    \Re\{\Sigma_R(\omega)\} \simeq -\frac{m_f g^2}{2\pi^2} \log \frac{\Lambda_b}{m_T^2} \omega 
\end{equation}
This indicates a renormalization of the quasiparticle weight $Z$ to,
\begin{equation}
    Z \simeq \left(1 + \frac{m_f g^2}{2\pi^2} \log\frac{\Lambda_b}{m_T^2} \right)^{-1}
\end{equation}

We now turn to the imaginary part of the fermion self-energy, which determines the decay rate.
To this end, we find the greater and lesser fermion self-energies, $\Sigma_{\rm mFL}^{>,<}$. The former is given as,
\begin{equation}
\begin{split}
    \Sigma_{\rm mFL}^>(k,\omega) &= \int_{q,q',\Omega} g^2 \Big (F^>(q,\Omega) + F^<(-q,-\Omega) \Big) G^>(k-q+q',\omega-\Omega) \\
    &= \int_{q,q',\omega'} 2 g^2 \big(b_0(\Omega) + 1 \big) A_F(q,\Omega) G^>(k-q+q',\omega-\Omega) \\
    & \simeq \frac{m_f g^2}{\pi} \int_{\Omega}  \big(b_0(\Omega) + 1 \big) \tan^{-1} \frac{\alpha \Omega}{m_T^2} \big(1-f_0(\omega-\Omega) \big) \,.
\end{split}
\end{equation}
Similarly, the latter is given as,
\begin{equation*}
\begin{split}
\Sigma_{\rm mFL}^< &= \int_{q,q',\Omega} g^2 \Big (F^>(q,\Omega) + F^<(-q,-\Omega) \Big) G^<(k-q+q',\omega-\Omega) \\
& \simeq \frac{m_f g^2}{\pi} \int_{\Omega}  \big(b_0(\Omega) + 1 \big) \tan^{-1} \frac{\alpha \Omega}{m_T^2} \big(f_0(\omega-\Omega) \big) \,.    
\end{split}
\end{equation*}
Combining these two results, $\Im\{\Sigma_{\rm mFL}^R\} = \frac{1}{2} \big\{\Sigma_{\rm mFL}^{>}+\Sigma_{\rm mFL}^{<} \big\}$, and we find,
\begin{equation}
\begin{split}
    \Im\{\Sigma_{\rm mFL}^R\} &= \frac{m_f g^2}{2\pi} \int_{\Omega}  \left\{b_0(\Omega) f_0(\omega-\Omega) + \big(b_0(\Omega) + 1 \big) \big(1-f_0(\omega-\Omega) \big) \right\} \tan^{-1} \frac{\alpha \Omega}{m_T^2} \,.
\end{split}
\end{equation}

\section{Quantum Boltzmann Equations}
\label{app:QBE}
We now provide details behind the quantum Boltzmann equations that we used in the main text to find the conductivity.
As we shall demonstrate, a major simplification in solving our quantum Boltzmann equation is that it is diagonal in frequency:
First, the static CDW disorder results in elastic scattering, and hence the collision integral for scattering off of the CDW disorder is diagonal in frequency.
On the other hand, the scattering off of critical bosons, albeit inelastic, relax a fermion's momentum instantaneously due to the uncorrelated nature of the Yukawa coupling; ergo, its collision integral is simply proportional to the scattering rate at that frequency and the density of excitations at that frequency and original momentum, and is diagonal in frequency.
An additional simplification comes from the fact that the fermion spectral function is sharply peaked around the Fermi surface -- this means that most excitations occur near the Fermi surface and we may 'integrate out' the momentum direction perpendicular to it \cite{KimLeeWen,NaveLee}.

Our starting point is given as \cite{mah00},
\begin{equation}
    A(\k,\omega)^2 \Im\{\Sigma_R(\k,\omega)\} \frac{e \k \cdot E}{m_f} {f_0}'(\omega) + \left(\frac{e \k \times B\hat{z}}{m_f} \right) \cdot \nabla_{\k} G^<(\k,\omega) = \Sigma^> G^< - \Sigma^< G^> \,.
    \label{eq:appQBE}
\end{equation}
At equilibrium $G^{<,>}$ satisfies the following relation,
\begin{equation*}
    G^{<}(\k,\omega) = f_0(\omega) A(\k,\omega) \,, \ \ G^{>}(\k,\omega) = \big( 1 - f_0(\omega) \big) A(\k,\omega) \,.
\end{equation*}
Once an electric field is applied, $G^{<,>}$ deviate from its equilibrium distribution. Let us call this deviation $\delta f(\k,\omega)$ so that,
\begin{equation}
    G^{<}(\k,\omega) = \big( f_0(\omega) + \delta f(\k,\omega) \big) A(\k,\omega) \,, \ \ G^{>}(\k,\omega) = \big( 1 - f_0(\omega) - \delta f(\k,\omega) \big) A(\k,\omega) \,.
\end{equation}

Let us determine the collision integral due to the interaction with the bosons.
To this end, we need to determine the non-equilibrium fermion self-energies.
The non-equilibrium interaction self-energy $\Sigma^>_{\rm mFL}$ is given as,
\begin{equation}
\begin{split}
    \Sigma_{\rm mFL}^>(\k,\omega) &= \int_{\q,\q',\Omega} g^2 \Big (F^>(\q,\Omega) + F^<(-\q,-\Omega) \Big) G^>(\k-\q+\q',\omega-\Omega) \\
    &= \int_{\q,\q',\Omega} 2 g^2 \big(b_0(\Omega) + 1 \big) A_F(\q,\Omega) G^>(\k-\q+\q',\omega-\Omega) \\
    & \simeq \frac{g^2}{\pi} \int_{\q',\Omega}  \big(b_0(\Omega) + 1 \big) \tan^{-1} \frac{\alpha \Omega}{m_T^2} \big(1-f_0(\omega-\Omega) - \delta f(\k-\q+\q', \omega-\Omega) \big) A(\k-\q+\q', \omega-\Omega) \\
    & \simeq \frac{g^2}{\pi} \int_{\Omega}  \big(b_0(\Omega) + 1 \big) \tan^{-1} \frac{\alpha \Omega}{m_T^2} \big(1-f_0(\omega-\Omega)\big) \,.
\end{split}
\end{equation}
where $A_F$ denotes the boson spectral function.
In the second line, we have made the assumption that the bosons are in thermal equilibrium.
This amounts to neglecting ``drag'' effects, by which the boson fluid is driven out of equilibrium when the fermions carry a current.
Because of the disordered nature of the interaction, fermion-boson scattering does not conserve momentum, and hence drag effects are expected to be weak and may be ignored.
Furthermore, in the fourth line, we have integrated over the momentum carried by the disorder:
The disorder can absorb any momentum $\q'$, and so the nonequilibrium portion of the integral simply becomes $\int_{\k'} \delta f(\k',\omega-\Omega) A(\k', \omega-\Omega)$ and vanishes since total charge is conserved.
Consequently, the interaction self-energy does not undergo a change, and the collision integral simply becomes,
\begin{equation}
    \Sigma_{\rm mFL}^> G^< - \Sigma_{\rm mFL}^< G^> = 2 \Im\{ \Sigma^R_{int} (\omega) \} A(\k, \omega) \delta f(\k, \omega)\,.
    \label{eq:ColIn_int}
\end{equation}
Ergo, the collision integral due to the interaction with the critical bosons is diagonal in frequency.

We now turn to the collision integral due to the collision with the CDW disorder.
Applying Eq.\eqref{eq:SigmaCDW} to the right hand side of \eqref{eq:appQBE}, it is given as,
\begin{equation}
\begin{split}
    \Sigma_{\rm dw}^> G^< - \Sigma_{\rm dw}^< G^> &= \int_{\k'} 2\pi J^2 \xi^2 e^{-\xi^2 |\k-\k' \pm \mathbf{Q}|^2} \Big( G^>(\k,\omega) G^<(\k',\omega) - G^<(\k,\omega) G^> (\k',\omega) \Big) \\
    &= \int_{\k'} 2\pi J^2 \xi^2 e^{-\xi^2 |\k - \k' \pm \Q|^2} A(\k,\omega) A(\k', \omega) \Big(\delta f(\k,\omega) - \delta f(\k',\omega) \Big) \,.
    \label{eq:ColIn_CDW}
\end{split}
\end{equation}
Note that this collision integral is also diagonal in frequency due to the elastic nature of the scattering process.

Now we make use of the fact that the fermion spectral function is sharply peaked around the Fermi surface and integrate out the momentum perpendicular to the Fermi surface.
We first perform a change of variables from $\k = (k_F + \Delta k) \hat{k}$ to $\hat{k}$ and $\Delta k = |\k|-k_F$, and define $f(\hat{k},\omega)$ as the generalized distribution function for fermions pointing in the $\hat{k}$ direction with frequency $\Omega$, given by,
$$f(\hat{k},\omega) = \int_{\Delta k} f(\k,\omega) = v_F \int_{\Delta k} G^< \big((k_F + \Delta k)\hat{k},\omega \big) \,.$$
Upon this change of variables, and integrating the left hand side of \eqref{eq:appQBE} with regards to $\Delta k$, we get, 
\begin{equation}
\begin{split}
    & \int_{\Delta k} [A(\k,\omega)]^2 \, \Im[\Sigma_R(\k,\omega)] \, \left( \frac{e \k \cdot \mathbf{E}}{m_f} \right) {f_0}'(\omega) + \left(\frac{e \k \times B\hat{z}}{m_f} \right) \cdot \nabla_{\k} G^<(\k,\omega) \\
    & \quad \simeq \int_{\Delta k} \frac{4 \Gamma^3} {\big((\omega - v_F \Delta k)^2 + \Gamma^2 \big)^2}
    e v_F\hat{k} \cdot \mathbf{E} {f_0}'(\omega) + \left(\frac{e \hat{k} \times B\hat{z}}{m_f} \right) \cdot \nabla_{\theta_k} G^<(\k,\omega) = e \hat{k} \cdot \mathbf{E} f_0'(\omega) + \frac{\omega_c}{v_F} \nabla_{\theta_k} \delta f(\hat{k},\omega)
    \label{eq:appQBE_RHS}
\end{split}
\end{equation}
Here, $\Gamma$ denotes the imaginary part of the self-energy, $\Im[\Sigma_R(\k,\omega) ]$.

Similarly, integrating the right handsides with regards to $\Delta k$, we find that the collision integral is given as,
\begin{equation}
    \int_{\Delta k} \Sigma^> G^< - \Sigma^< G^> = \frac{2}{v_F}\Im\{\Sigma_{\rm mFL}^R(\omega)\} \delta f(\hat{k},\omega) + \int_{\hat{k'}} \frac{2\pi J^2 \xi^2 k_F}{v_F^2} e^{-\xi^2 |k_F \hat{k}-k_F \hat{k'} \pm \Q|^2} \Big(\delta f(\hat{k'},\omega) - \delta f(\hat{k},\omega) \Big)
    \label{eq:appQBE_LHS}
\end{equation}

Now, we define $\delta f(\hat{k},\omega) = f_0(\omega) \big(1 - f_0(\omega)\big) g(\hat{k},\omega)$.
Rewriting Eq.\eqref{eq:appQBE} in terms of $g$ and dividing both sides by $f_0(\omega) \big(1-f_0(\omega)\big)$ simplifies to,
\begin{equation}
    e \beta v_F \hat{k} \cdot \mathbf{E} + \omega_{c0} \nabla_{\theta_k} g(\hat{k},\omega) = 2 \Im\{\Sigma_{\rm mFL}^R(\omega)\} g(\hat{k},\omega) + \int_{\hat{k'}} \frac{4\pi J^2 (k_F\xi)^2 }{\varepsilon_F} e^{-\xi^2 |k_F \hat{k}-k_F \hat{k'} \pm \Q|^2} \Big(g(\hat{k'},\omega) - g(\hat{k},\omega) \Big)
\end{equation}
where $\omega_{c0} = \frac{eB}{m_f}$ denote the bare cyclotron frequency.

Let us define $\tilde g(\hat{k},\tilde\omega = \beta\omega) = \frac{g(\hat{k},\omega)}{e v_F \beta^2 E}$.
This rescaled distribution function $\tilde g$ is much smoother than $\delta f$ \cite{KimLeeWen,NaveLee}, allowing for greater accuracy in numerically solving the quantum Boltzmann equations.
The quantum Boltzmann equations for $\tilde g$ is given as,
\begin{equation}
\begin{split}
    \cos \theta_k - \frac{\omega_{c0}}{T} \partial_{\theta_k} \tilde g(\hat{k},\tilde\omega) = 2 \beta \Im\{ \Sigma_{\rm mFL}(\omega) \} \tilde g(\hat{k},\tilde\omega) + \frac{4\pi J^2} {\varepsilon_F T} \int_{\hat{k'}} (k_F\xi)^2 e^{-\xi^2 |k_F \hat{k}-k_F \hat{k'} \pm \Q|^2} \Big(\tilde g(\hat{k'},\tilde\omega) - \tilde g(\hat{k},\tilde\omega) \Big)
    \label{eq:BEf_rescaled}
\end{split}
\end{equation}
We numerically find the rescaled distribution function $\tilde g$ by discretizing the angle around the Fermi surface and the frequency $\omega$.
After finding the rescaled distribution $\tilde g$ by solving Eq.\eqref{eq:BEf_rescaled}, we find the field induced current $J$ and subsequently the conductivity.
Assuming that the electric field is pointing in the $x$ direction without loss of generality, the charge conductance is given as,
\begin{equation}
\begin{split}
    \begin{pmatrix}
    \sigma^{xx} \\
    \sigma^{xy}
    \end{pmatrix}
    &= \frac{4 e^2 \varepsilon_F}{T} \int_{\hat{k},\tilde\omega} \begin{pmatrix}
    \cos \theta_k \\
    \sin \theta_k
    \end{pmatrix} \sech^2 \frac{\beta\omega}{2} \tilde{g}(\hat{k},\tilde\omega) \,.
    \label{eq:cond}
\end{split}
\end{equation}    
The resistance can then be extracted from Eq.\eqref{eq:cond} through the relation, $R_{xx} = \sigma_{xx}/(\sigma_{xx}^2 + \sigma_{xy}^2)$.

Finally, we note that from \eqref{eq:BEf_rescaled} we may understand the emergence of the Kohler scaling.
With the first term on the right-hand side of the rescaled quantum Boltzmann equations, $\beta \Im\{\Sigma\}$ is a function of $\tilde \omega$ and independent of temperature. Similarly, with the second term, the coefficient $4\pi J^2 / \varepsilon_F T$ is effectively infinite at the low-temperature limit that we work with.
This means that $\tilde g$ is a function of $\omega_{c0}/T$, i.e. $B/T$.
Hence, applying this result to $\eqref{eq:cond}$, we find that $\sigma^{xx,xy}/T$ is a function of $B/T$, resulting in the Kohler-like scaling of magnetoresistance.


\vfill

\end{document}